\PassOptionsToPackage{unicode}{hyperref}
\PassOptionsToPackage{hyphens}{url}

\documentclass[11pt]{article}
\usepackage[a4paper,margin=1in]{geometry}

\usepackage{caption}
\usepackage{xcolor}
\usepackage{amsmath,amssymb}
\usepackage{iftex}
\ifPDFTeX
  \usepackage[T1]{fontenc}
  \usepackage[utf8]{inputenc}
  \usepackage{textcomp} 
\else 
  \usepackage{unicode-math} 
  \defaultfontfeatures{Scale=MatchLowercase}
  \defaultfontfeatures[\rmfamily]{Ligatures=TeX,Scale=1}
\fi
\usepackage{lmodern}
\ifPDFTeX\else
\fi
\IfFileExists{upquote.sty}{\usepackage{upquote}}{}
\IfFileExists{microtype.sty}{
  \usepackage[]{microtype}
  \UseMicrotypeSet[protrusion]{basicmath} 
}{}
\makeatletter
\@ifundefined{KOMAClassName}{
  \IfFileExists{parskip.sty}{%
    \usepackage{parskip}
  }{
    \setlength{\parindent}{0pt}
    \setlength{\parskip}{6pt plus 2pt minus 1pt}}
}{
  \KOMAoptions{parskip=half}}
\makeatother
\usepackage{longtable,booktabs,array}
\usepackage{calc} 
\usepackage{etoolbox}
\makeatletter
\patchcmd\longtable{\par}{\if@noskipsec\mbox{}\fi\par}{}{}
\makeatother
\IfFileExists{footnotehyper.sty}{\usepackage{footnotehyper}}{\usepackage{footnote}}
\makesavenoteenv{longtable}
\usepackage{graphicx}
\makeatletter
\newsavebox\pandoc@box
\newcommand*\pandocbounded[1]{
  \sbox\pandoc@box{#1}%
  \Gscale@div\@tempa{\textheight}{\dimexpr\ht\pandoc@box+\dp\pandoc@box\relax}%
  \Gscale@div\@tempb{\linewidth}{\wd\pandoc@box}%
  \ifdim\@tempb\p@<\@tempa\p@\let\@tempa\@tempb\fi
  \ifdim\@tempa\p@<\p@\scalebox{\@tempa}{\usebox\pandoc@box}%
  \else\usebox{\pandoc@box}%
  \fi%
}
\def\fps@figure{htbp}
\makeatother
\usepackage{bookmark}
\IfFileExists{xurl.sty}{\usepackage{xurl}}{} 
\hypersetup{
  hidelinks,
  pdfcreator={LaTeX via pandoc}}

\author{}
\date{}

\usepackage{booktabs}
\usepackage{siunitx}
\DeclareSIUnit\year{yr}

\title{The Frequency of Solar Eclipses for a Given Place: A New Approach to a Classic Question}

\author{}

\date{}

\begin{document}
\maketitle

\maketitle
\vspace{-2.5em}

\begin{center}
Graham Jones, Renate Mauland-Hus, Frank Thomas Tveter, Anne Buckle, S\'ebastien Emonet, Adalbert Michelic, Anna Smith, David Usken, Steffen Thorsen

\vspace{0.6em}

\small
timeanddate.com, Stavanger, Norway\\
\href{mailto:graham@timeanddate.com}{graham@timeanddate.com}

\vspace{0.8em}

\textit{Accepted for publication in the \emph{Journal of the British Astronomical Association}.}
\end{center}

\vspace{1.2em}

\section*{Abstract}

In a classic 1982 paper in this journal, Jean Meeus used a statistical
approach for finding the mean frequency of a total and an annular
eclipse of the Sun at a given place on the surface of the Earth. In this
current paper we tackle the problem again, taking advantage of the much
greater computing power now available. We obtain narrower estimates of
once every 373 ± 7 years for a total eclipse, and once every 226 ± 4
years for an annular eclipse. In addition, we obtain a result of once
every 2.59 ± 0.02 years for a partial eclipse. We provide further
insights into the ``latitude effect'', showing that solar eclipses occur
most frequently around the Arctic and Antarctic Circles. We also show
how the gradual shift of aphelion and perihelion with respect to the
seasons produces a \textasciitilde21,000-year cycle in the frequency of
eclipses in the Northern and Southern Hemispheres.

\section{Introduction}\label{introduction}

In 1982, writing in this journal, astronomer and mathematician Jean
Meeus asked the following question: ``How often can a total or an
annular eclipse of the Sun be expected at a given point on the Earth's
surface?''.{[}1{]} Previously, in their 1926 textbook ``Astronomy'',
Russell, Dugan, and Stewart had stated that ``we find that in the long
run a total eclipse happens at any given station only once in about 360
years'', but without providing any details.{[}2{]}

Meeus attacked the problem statistically, calculating the local
circumstances at 408 ``standard points'' on the Earth's surface for
every solar eclipse in the period 1700--2299 Common Era (CE). A period
of 600 years was chosen to avoid any effect from a known 586-year cycle
in the frequency of lunar eclipse tetrads.{[}3{]} Meeus arrived at the
following mean frequencies for any given point: once every 375 ± 16
years for a total eclipse, and once every 224 ± 7 years for an annular
eclipse. He also identified a ``latitude effect'' where the mean
frequency of a total eclipse is higher in the Northern Hemisphere than
the Southern Hemisphere, while the reverse is true for an annular
eclipse.

Meeus was limited by the computing power then available: he used an
HP-85, a small personal computer that was introduced in 1980.{[}4{]} One
potential issue is revealed by an inspection of the computer outprint
sample, from 1973 June 30 to 1983 June 11, that Meeus included in his
paper: the computer appears to have missed an annular eclipse on 1977
April 18 at the standard point with latitude 10°S and longitude
30°E.{[}a{]}{[}5{]}

Some four decades after Meeus's paper was published, we decided to
revisit the question of eclipse frequency. The much greater computing
power now available would enable us to use highly precise areas instead
of reference points, over a longer period of time; it would also enable
us to include partial solar eclipses in our calculations. Questions
about eclipse frequencies are not simply of technical interest: they are
an important part of the general public's fascination with solar
eclipses. Meeus's figure of ``once every 375 years, on average'' is
often quoted in mainstream media reports, for example, as during the
buildup to the total solar eclipse across North America in 2024.{[}6{]}

\section{Methods}\label{methods}

Instead of taking a 600-year period, as Meeus did, we took a period of
14,999 years, from 0001 January 1 to 14999 December 31. We used the
Gregorian calendar throughout this period; in other words, we did not
switch to the Julian calendar for dates before 1582 October 15.
Furthermore, instead of considering a set of 408 points, as Meeus did,
we divided the Earth's surface into 180 latitude bands. Each band has a
height of 1° of latitude, and a width of 360° of longitude. For example,
the latitude band that includes London in the UK, at a latitude of
51.5°N, is effectively a four-sided polygon with its four corners at
51°N, 180°W; 51°N, 180°E; 52°N, 180°E; and 52°N, 180°W.

For each solar eclipse within the given period, we determined the path
of the Moon's shadow across the Earth's surface. Instead of calculating
Besselian elements, we used a search algorithm to find the edge of the
shadow. Our algorithm distinguished between three different parts of the
shadow: the penumbra, which produces a partial eclipse; the umbra, which
produces a total eclipse; and the antumbra, which produces an annular
eclipse. For each eclipse, we determined what fraction of each latitude
band is covered by each part of the Moon's shadow. From this, we
determined the average number of eclipses at a given place within the
latitude band, which in turn gave us the average frequency.

As an illustration of the method we used, suppose that, worldwide, there
are four partial solar eclipses in a period of two years. Let us
consider a particular latitude band near one of the poles, and say there
are three different cases: (a) at two of the eclipses, the Moon's shadow
covers no part of the band; (b) at one eclipse, the shadow covers half
the band; (c) at the other eclipse, the shadow covers all of the band.
For each event, the probability that a randomly chosen point within the
band will see the eclipse is equal to the area fraction of the band
covered by the Moon's shadow. In our three cases, these probabilities
are (a) 0, (b) 0.5, and (c) 1. On average, how many eclipses would we
expect our random point in the band to see over the two-year period? It
would be the sum of the area fractions, or 0+0+0.5+1=1.5. Therefore, the
average frequency per year for a randomly chosen point is 1.5÷2=0.75. In
this exercise, we have assumed nothing about the latitude of the random
point, other than that it is somewhere within the latitude band.
Furthermore, we have assumed nothing about the longitude of the random
point: the geographic longitude of an eclipse depends on where Earth is
in its roughly 24-hour rotational period, which has no connection with
the timing of eclipses.

Finally, to calculate the average frequency for a given place anywhere
on the Earth's surface, we weighted each latitude band according to its
size as a fraction of Earth's total surface area. This final step
accounted for latitude bands being larger around the equator, and
smaller around the poles.

We used our own algorithms, together with JPL's planetary and lunar
ephemeris DE431, the U.S. Naval Observatory's source-code library NOVAS,
and the IAU's SOFA software. We ran 30 parallel jobs continuously for
102 days on two Supermicro servers. Over the runtime, the workload
consumed around 147,000 core hours and 662,000\,GiB hours of memory.

\section{Results}\label{results}

A sample of the computer output for five latitude bands is shown in
table 1; this sample data covers the period 2000 January 1 to 2999
December 31. Globally, there are 2,384 solar eclipses within this span
of 1,000 years. The data for all latitude bands, for the same period, is
included in the appendix as table A1. The appendix also includes a link
to our data for the entire span of 14,999 years. An explanation of the
column headings in table 1 and table A1 is as follows.

\begin{itemize}
\item
  Band midpoint --- the north-south midpoint of the latitude band in
  degrees. For example, 51.5 is the midpoint of the latitude band
  51°N--52°N
\item
  Area --- the size of the latitude band as a fraction of the total
  surface area of the Earth, where the total surface area is normalised
  to 1
\item
  Count: All --- the number of eclipses, of any type, occurring anywhere
  within the latitude band
\item
  Average: All --- the average number of eclipses, of any type, for a
  given place within the latitude band
\item
  Count: Totality --- the number of total eclipses occurring anywhere
  within the latitude band
\item
  Average: Totality --- the average number of total eclipses for a given
  place within the latitude band
\item
  Count: Annularity --- the number of annular eclipses occurring
  anywhere within the latitude band
\item
  Average: Annularity --- the average number of annular eclipses for a
  given place within the latitude band
\end{itemize}

\begin{table}[htbp]
\centering
\small
\resizebox{\linewidth}{!}{%
\begin{tabular}{
  S[table-format=2.1]
  S[table-format=1.6]
  S[table-format=4.0]
  S[table-format=3.2]
  S[table-format=3.0]
  S[table-format=1.2]
  S[table-format=3.0]
  S[table-format=1.2]
}
\toprule
{Band midpoint} & {Area} &
{\shortstack{Count:\\All}} & {\shortstack{Average:\\All}} &
{\shortstack{Count:\\Totality}} & {\shortstack{Average:\\Totality}} &
{\shortstack{Count:\\Annularity}} & {\shortstack{Average:\\Annularity}} \\
\midrule
50.5 & 0.005570 & 1175 & 414.33 & 154 & 2.88 & 192 & 4.71 \\
51.5 & 0.005453 & 1168 & 416.89 & 147 & 3.11 & 181 & 4.40 \\
52.5 & 0.005333 & 1159 & 419.59 & 146 & 3.17 & 182 & 4.45 \\
53.5 & 0.005213 & 1155 & 422.02 & 141 & 3.08 & 181 & 4.54 \\
54.5 & 0.005090 & 1147 & 424.14 & 138 & 3.12 & 179 & 4.68 \\
\bottomrule
\end{tabular}
}
\caption{Sample of our computer output for 2000--2999 CE. The column
headings are explained above.}
\end{table}

From the computer output for the entire span of 14,999 years, we were
able to determine the average return periods of eclipses at any given
place on the Earth's surface. The return period is the number of years
between successive eclipses; it is the inverse of frequency. We were
able to narrow Meeus's estimates for the return periods of total and
annular eclipses; we were also able to obtain an estimate for the return
period of partial eclipses, including partial eclipses that become total
or annular at some stage. A comparison between Meeus's results and our
results is shown in table 2.

\begin{table}[htbp]
\centering
\small
\begin{tabular}{
  l
  S[table-format=3.0] @{${}\pm{}$} S[table-format=2.0]
  S[table-format=3.0] @{${}\pm{}$} S[table-format=1.0]
  l
}
\toprule
Study &
\multicolumn{2}{c}{Total} &
\multicolumn{2}{c}{Annular} &
{Partial} \\
\midrule
Meeus        & 375 & 16 & 224 & 7 & Not given \\
Our results  & 373 & 7  & 226 & 4 & $2.59 \pm 0.02$ \\
\bottomrule
\end{tabular}
\caption{Average return periods of solar eclipses at any given place, in
years.}
\end{table}

We also obtained average return periods at different latitudes. For
example, using the figures in table 1 as an illustration, for a given
place at the latitude of London, the ``Average: All'' column tells there
are, on average, around 417 eclipses within this span of 1,000 years:
this leads to an average return period of 2.40 years. Figure 1 shows the
average return period of eclipses of any type by latitude, based on the
computer output for the entire span of 14,999 years.

\begin{figure}[htbp]
  \centering
  \includegraphics[width=\linewidth]{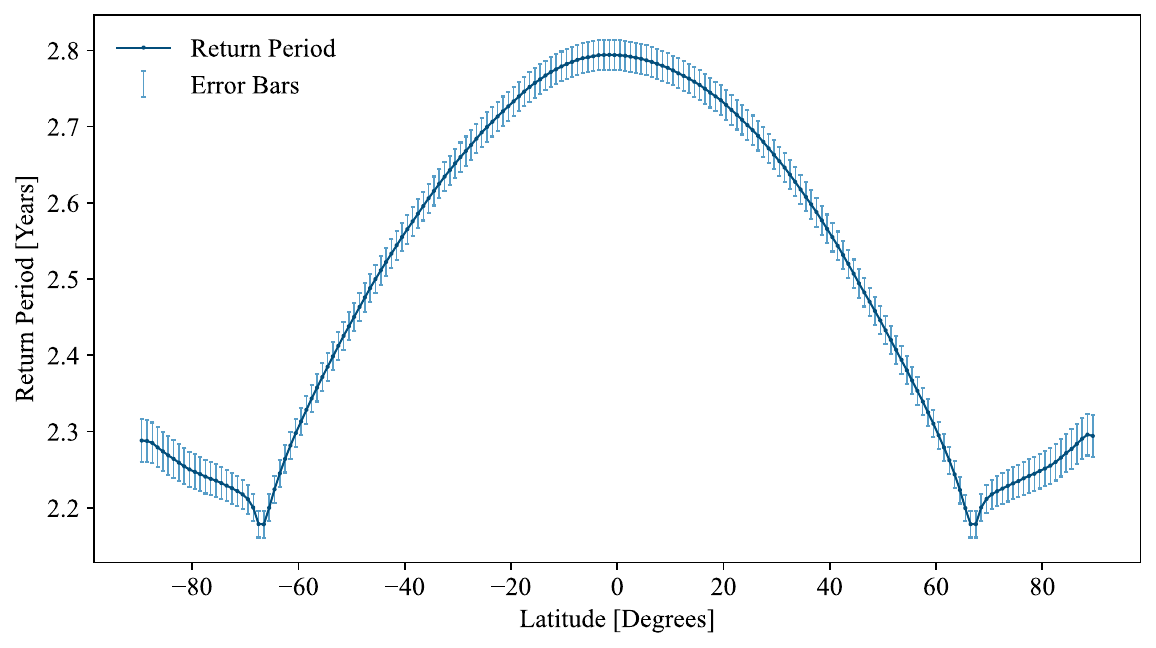}
  \caption{Average return period of solar eclipses of any type, by latitude. The error bars indicate the confidence interval, applying the central limit theorem.}
  \label{fig:eclipse-return-period}
\end{figure}

In addition, we obtained results for the number of eclipses by latitude
in different millennia.{[}b{]} For comparison purposes, we converted
this number to a value we call the ``Meeus frequency'': this is the
overall number of eclipses we would expect to see at a set of 24 given
points within a particular latitude band, over a 600-year period within
a particular millennium; it is equivalent to the number of eclipses that
Meeus counted at 24 points along a line of latitude over 600
years.{[}c{]} The Meeus frequency \(f_{Meeus}\) is given by the
following formula, where \(N_{avg}\) is the average number of eclipses
for a given place within a latitude band, as shown in the ``Average''
columns in table 1 and table A1.

\[
f_{\text{Meeus}}
  = 24\,N_{\text{avg}}
    \cdot
    \frac{\qty{600}{\year}}{\qty{1000}{\year}}
\]

For example, within our latitude band 51°N--52°N for 2000--2999 CE, the
average number of total eclipses for any given place is around 3.11;
this leads to a Meeus frequency of 44.8, given by 24×3.11×600÷1000. This
matches reasonably well with the third-degree polynomial Meeus fitted to
his chart showing the number of eclipses at his standard points for
1700--2299 CE, which indicates a value of 46.1 at 51.5°N.

Figure 2 shows the Meeus frequency of total eclipses within each of our
180 latitude bands, in each of our 15 millennia. Each value is shown as
a point; there are 2,700 points overall. Each millennium has been given
its own colour, ranging from light blue for the 1st millennium
(0001--0999 CE) to dark blue for the 15th millennium (14000--14999 CE).
The higher the Meeus frequency, the higher the number of total eclipses
at that latitude within that millennium.

\begin{figure}[htbp]
  \centering
  \includegraphics[width=\linewidth]{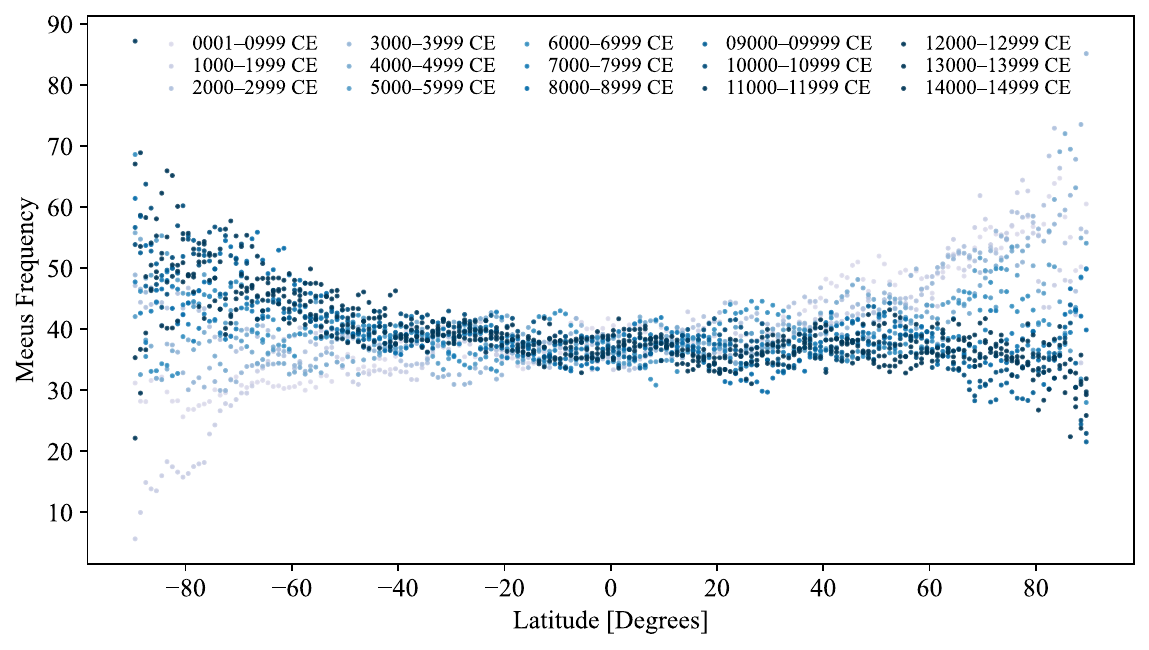}
  \caption{Meeus frequency of total solar eclipses, by latitude and by
millennium.}
  \label{fig:meeus-frequency}
\end{figure}

Figure 3 shows the same data for total eclipses as figure 2, using the
same colours, but each millennium is shown as a separate chart that
includes a confidence interval, a trend line using the least squares
linear regression method, and a label giving the slope of the trend
line.

\begin{figure}[htbp]
  \centering
  \includegraphics[width=\linewidth]{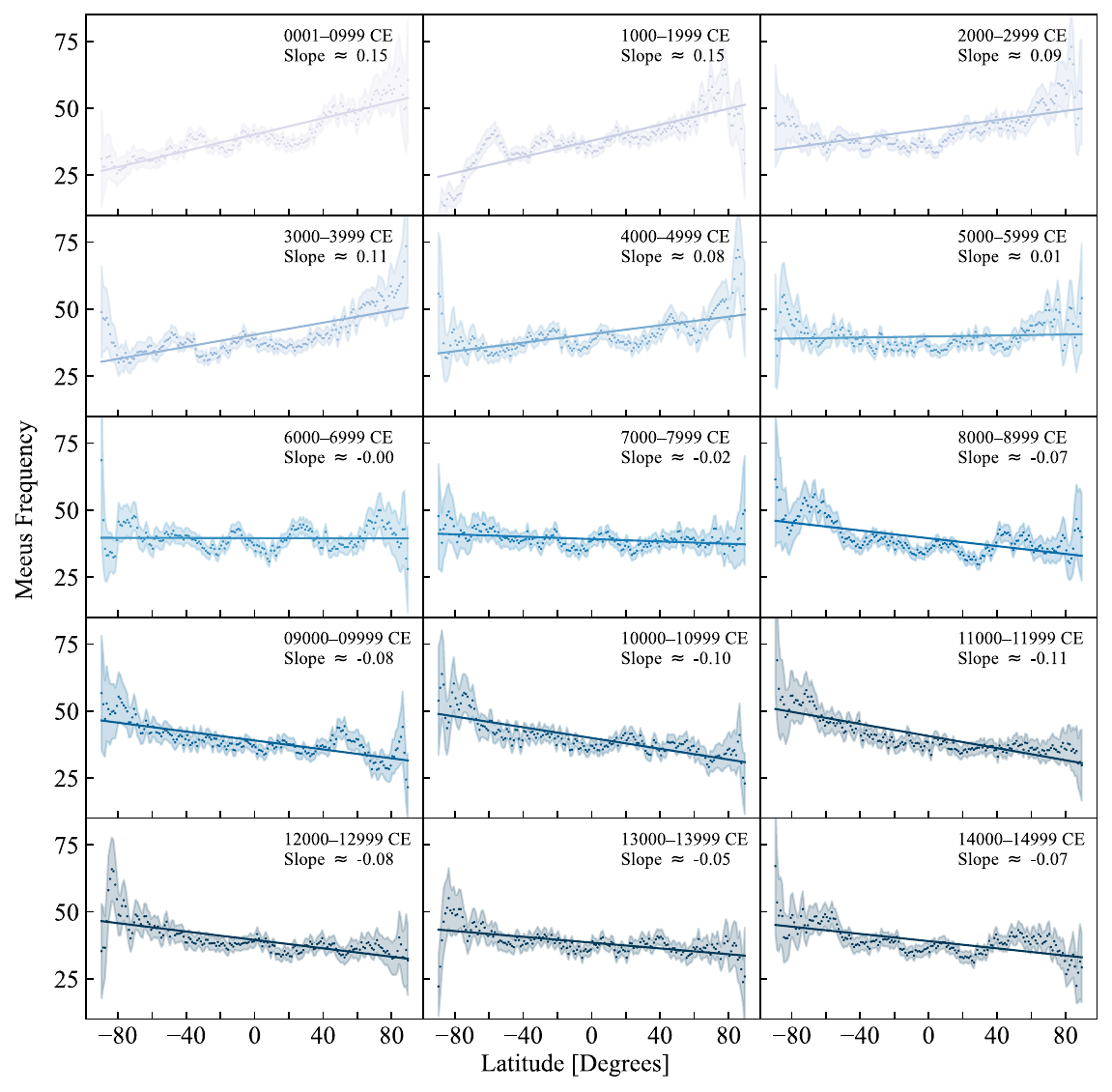}
  \caption{Fifteen charts, one for each millennium, showing the Meeus frequency
of total solar eclipses, by latitude, with trend lines.}
  \label{fig:millennia-charts}
\end{figure}

Figure 4 is the equivalent of figure 2 for annular eclipses, while
figure 5 is the equivalent of figure 2 for partial eclipses,
including the partial phases of total and annular eclipses. In the case
of partial eclipses, a comparison with Meeus's results is not possible,
because he only considered total and annular eclipses.

\begin{figure}[htbp]
  \centering
  \includegraphics[width=\linewidth]{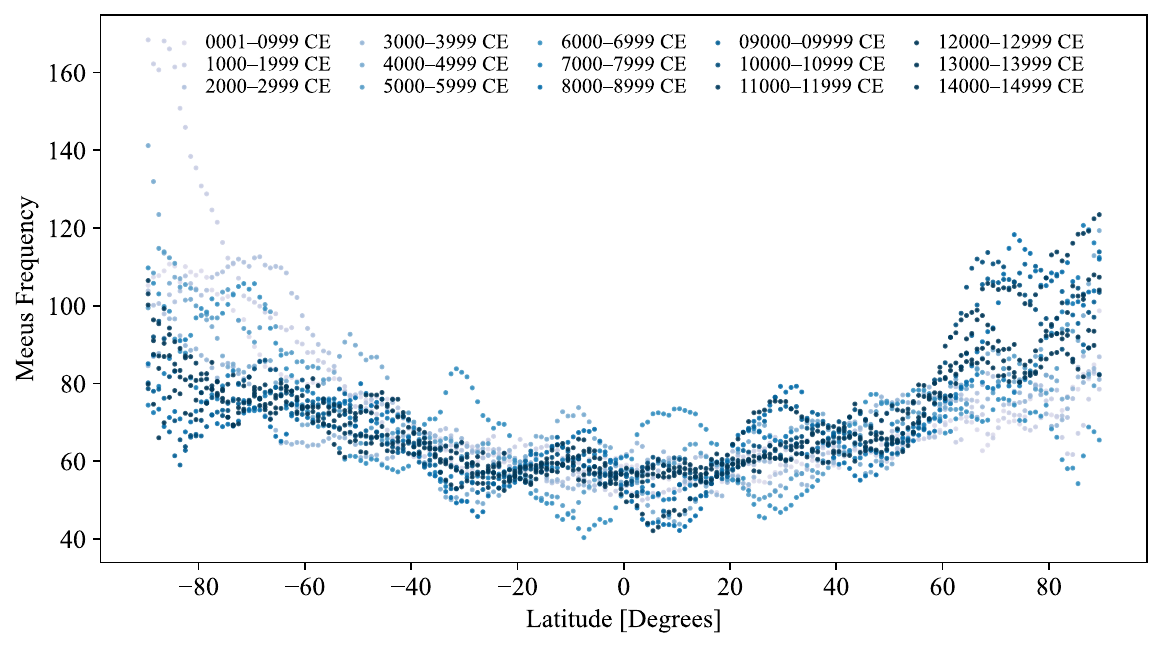}
  \caption{Meeus frequency of annular solar eclipses, by latitude and by
millennium.}
  \label{fig:annular-eclipses}
\end{figure}

\begin{figure}[htbp]
  \centering
  \includegraphics[width=\linewidth]{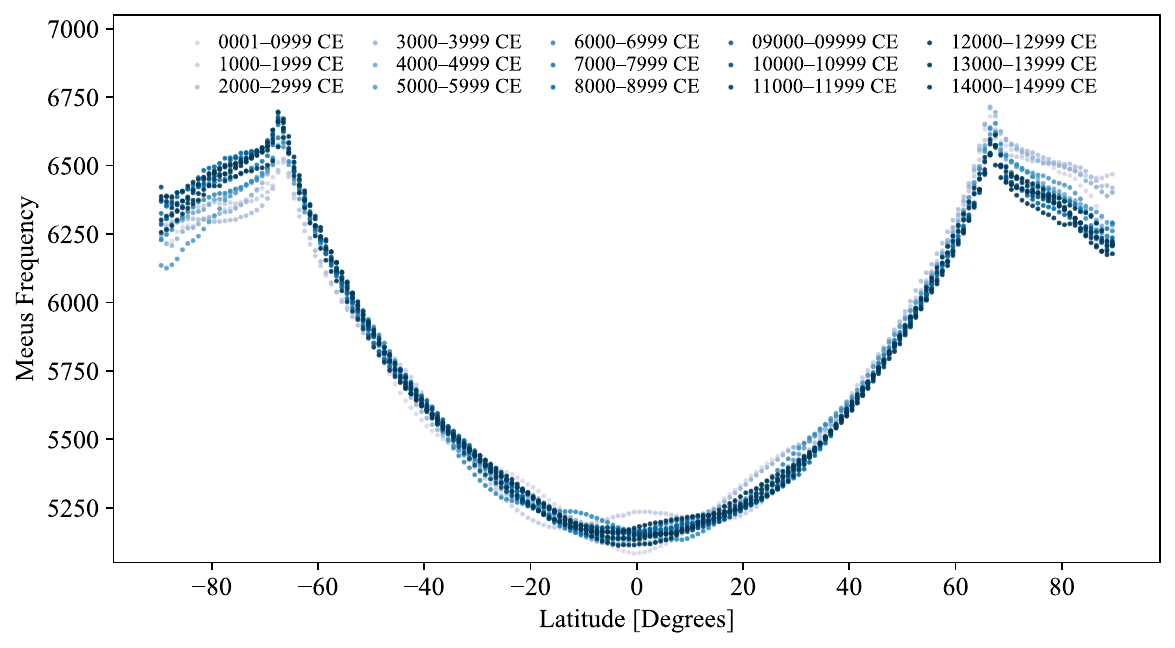}
  \caption{Meeus frequency of solar eclipses of any type, by latitude and by
millennium.}
  \label{fig:eclipses-any-type}
\end{figure}

\section{Discussion}\label{discussion}

Our results for the average return periods of total and annular eclipses
at any given place support Meeus's findings from 1982. We have narrowed
the uncertainty, and we suggest a small shift in the nominal values: 373
± 7 years for total eclipses, compared to Meeus's 375 ± 16 years; and
226 ± 4 years for annular eclipses, compared to 224 ± 7 years. The
closeness of the two sets of results supports Meeus's choice of 408
reference points and a 600-year search period. Nevertheless, in many
ways, it is remarkable that Meeus was able to obtain such results with,
relatively speaking, such limited computing power. We mentioned in our
introduction that the computer output sample Meeus included in his paper
appears to miss an annular eclipse at one of his standard points. There
is also a potential query about some of his other data. When we took
Meeus's figures for the mean time intervals between eclipses, as shown
in the fourth and fifth columns of table 1 in his 1982 paper, and
weighted them in the way he described in the paper's appendix, we
obtained a different outcome: mean frequencies of once in 382 years for
total eclipses, and once in 233 years for annular eclipses.

The average return period of partial eclipses at any given place,
including the partial phases of total and annular eclipses, is a new
finding: 2.59 ± 0.02 years. Globally, over our entire span of 14,999
years, there are 35,538 solar eclipses; dividing 14,999 by 35,538 gives
us a global return period of around 0.42 years. Broadly speaking, if we
divide 0.42 by 2.59, we can say that every time a solar eclipse occurs
somewhere in the world, there is a roughly one-in-six chance that any
given point on Earth's surface will be included. Another way of
expressing this is to say there is a roughly one-in-six chance that any
given town or city will be included. However, these local return periods
vary with latitude.

Our results provide further insights into the latitude effect: figure 1
shows that the frequency of a solar eclipse of any type is highest
around the Arctic and Antarctic Circles, at latitudes of 66.6°N and
66.6°S respectively, and lowest around the equator. Why is this? Around
the equator, the Sun tends to rise and set straight upwards and
downwards from the horizon. Away from the equator, the Sun rises and
sets at more of a slope, which increases the amount of time that some
part of the Sun is above the horizon: the leading edge at sunrise, the
trailing edge at sunset. This effect is most pronounced at the Arctic
and Antarctic Circles, where the Sun spends the most time skimming along
the horizon. One of the conditions for a solar eclipse to occur at a
given location is that some part of the Sun must be above the horizon:
consequently, the more time that some part of the Sun is above the
horizon, the more opportunity there is for an eclipse to occur.

The Sun also skims along the horizon at the North and South Poles. Here,
however, the skimming occurs around the equinoxes, whereas at the Arctic
and Antarctic Circles the skimming occurs around the solstices. Around
the equinoxes, from one day to the next, the Sun's path across the sky
changes more quickly than it does around the solstices. Therefore, over
the course of a year, the Sun spends less time above the horizon at the
North Pole than it does at the Arctic Circle; the same is true for the
South Pole and the Antarctic Circle.

Again, we are revisiting the work of Meeus, who showed in a separate
study, published in 2010, that solar eclipse frequency peaks around the
polar circles.{[}7{]} Here, however, Meeus ignored the effects of
atmospheric refraction and the size of the solar disk: he treated the
Sun as being above the horizon only when the geometric altitude of the
centre of the Sun's disk is greater than zero. In the same study, Meeus
further noted that for eclipses where the Sun is above the horizon at
the maximum phase, the decrease in frequency between the polar circles
and the poles does not occur. The peak at the polar circles, therefore,
is provided by eclipses where the Sun either rises after or sets before
the maximum phase of the eclipse. Given that he was ignoring refraction
and the Sun's diameter, Meeus observed that the overall decrease in
eclipse frequency between the polar circles and the poles coincides with
a decrease in the number of sunrises and sunsets. Certainly, the
dependence on latitude is complex: in another separate study, published
in 2004, Meeus showed that for partial eclipses where the Moon's umbral
and antumbral shadow miss the Earth completely, the point of greatest
eclipse lies within one of two latitude bands that extend 5.5° either
side of the polar circles.{[}8{]} This aspect of eclipses may also have
a bearing on the spikes at the polar circles we can see in figure 1
above.

For total eclipses, Meeus showed in his 1982 paper that their frequency
is higher in the Northern Hemisphere than the Southern Hemisphere, while
the reverse is true for annular eclipses. As Meeus noted, the Sun is
above the horizon the longest during the summer. For the Northern
Hemisphere, summer coincides with aphelion, in early July, when Earth is
furthest from the Sun and the apparent size of the Sun's disk is
smallest. The smaller the Sun's disk, the greater the chance of an
eclipse being total rather than annular. Meanwhile, for the Southern
Hemisphere, summer coincides with perihelion, in early January, when
Earth is closest to the Sun and the chance of an eclipse being total is
less.

With our results, we were able to go one step further and show how the
frequency of total and annular eclipses in each hemisphere evolves over
time. Meeus studied the present day: he chose a period that includes the
end of the 2nd millennium CE and the beginning of the 3rd millennium CE.
In figure 2, which shows the frequency of total eclipses by latitude and
by millennium, the 2nd and 3rd millennia are in light blue colours; for
these points, in general, the frequency does indeed increase as latitude
increases from -90° to +90°. However, in our study, we were able to
include future millennia, up to and including the 15th millennium CE. In
figure 2, these future millennia are in dark blue colours; for these
points, frequency generally \emph{decreases} as latitude increases.

In figure 3, the data points have been rearranged to show the overall
pattern more clearly. In the present day, the slope of the line is
positive: in other words, total eclipses are more frequent in the
Northern Hemisphere. In future millennia, the slope of the line is
negative, ie, total eclipses are more frequent in the Southern
Hemisphere. In the 7th millennium CE, the slope of the line is roughly zero,
meaning the frequency of total eclipses is the same for both
hemispheres.

The pattern shown in figure 3 can be explained in the following way. The
dates of aphelion and perihelion drift through the calendar in a cycle
lasting around 21,000 years.{[}9{]} Around 1,000 years ago, aphelion
roughly coincided with the June solstice, giving maximum advantage to
the Northern Hemisphere in terms of how long the Sun is above the
horizon when it is at its smallest apparent size, and the chance of an
eclipse being total rather than annular is greatest. In the future,
around 9,500 years from now, aphelion will roughly coincide with the
December solstice, reversing the situation. Midway between these times,
aphelion will roughly coincide with the September equinox, when the
amount of daylight is the same at all latitudes, giving neither
hemisphere an advantage with regard to the apparent size of the Sun.

Over the same timespan, the date of perihelion, which favours annular
eclipses, drifts from around the December solstice to around the June
solstice. This is reflected in figure 4: comparing the light blue dots
for the present day with the dark blue dots for future millennia, the
frequency of annular eclipses decreases in the Southern Hemisphere, and
increases in the Northern Hemisphere. However, the pattern for annular
eclipses shown in figure 4 is not simply the reverse of the pattern for
total eclipses shown in figure 2. For total eclipses, the lowest Meeus
frequencies tend to occur around one of the poles; for annular eclipses,
they tend to occur around the equator. As Meeus noted in his paper,
Earth's equatorial regions are closer to the Moon, which increases the
apparent size of the Moon's disk, and decreases the chance of an annular
eclipse. Meeus also noted that the Moon's shadow moves across the
Earth's surface in approximately the same direction as the Earth
rotates: this slows the speed at which the Moon's shadow moves across a
map from west to east. This effect, which increases the duration of an
eclipse for a given place whilst reducing its frequency, is greatest
around the equator.

Overall, the roughly 21,000-year cycle in the alignment of aphelion and
perihelion with the seasons affects the frequency of eclipses of any
type, as shown in figure 5. In the present day, summer lasts longer in
the Northern Hemisphere than the Southern Hemisphere; over the course of
a year, the Sun spends more time above the horizon in the Northern
Hemisphere, meaning there is more opportunity for an eclipse to occur.
In future millennia, this pattern is reversed.

Over the 14,999-year period of our study, Earth's eccentricity
decreases, giving us less extreme aphelions and perihelions; we might
expect this to lead to less extreme differences between the number of
eclipses in each hemisphere. Whilst we did not investigate this in
detail as part of our study, we can remark that, in figure 3, the change
in the steepness of the line is not symmetrical either side of the 7th
millennium CE, when aphelion and perihelion coincide with the equinoxes.
Approximately five millennia before this, when Earth's eccentricity is
greater, the maximum steepness of the line is 0.15; five millennia or so
afterwards, when Earth's eccentricity is less, the maximum steepness of
the line is -0.11. In other words, at first glance, the difference
between the Northern and Southern Hemispheres does indeed appear to
lessen. However, this needs further investigation.

It is important to stress, as Meeus did, that the return periods we
obtained are the \emph{average} time intervals between solar eclipses.
Although eclipses follow predictable cycles at a global level, such as
the Saros cycle, these cycles do not apply to a particular location. As
Meeus commented, total and annular eclipses ``take place at very
irregular intervals for a given place.'' We can give a simple
illustration of how this applies to partial eclipses, too, by taking the
next five upcoming eclipses in London, following the 31\% partial
eclipse on 2025 March 29, as shown in table 3. The return period for
these five eclipses is 2.28 years; in other words, the average interval
for London quickly closes in on the figure of 2.40 years that we gave in
our results above. However, the \emph{actual} intervals from one eclipse
to the next are very irregular, ranging from 0.49 years to 6.22 years.

\begin{table}[htbp]
\centering
\small
\begin{tabular}{
  l
  S[table-format=2.0]
  S[table-format=1.2]
}
\toprule
Date & {Obscuration (\%)} & {Interval} \\
\midrule
2026 August 12 & 91 & 1.37 \\
2027 August 2  & 42 & 0.97 \\
2028 January 26 & 50 & 0.49 \\
2030 June 1    & 48 & 2.35 \\
2036 August 21 & 60 & 6.22 \\
\bottomrule
\end{tabular}
\caption{Next five solar eclipses in London, including the maximum
obscuration of the Sun's disk as a percentage, and the interval since
the previous eclipse in years.{[}10{]}}
\end{table}

There are a number of small uncertainties in our results. We did not
take the elevation of the Earth's surface into account; all our eclipse
calculations are based on sea level. We used approximations for the size
and shape of the Sun and the Moon, and we ignored the fact that the Moon is
not a perfect sphere: it has both an uneven shape, and a rough,
mountainous edge, which, in turn, gives its shadow an uneven shape. In
addition, there are uncertainties in the ephemeris we used, which tend
to increase as we move further away from the present day. On the other
hand, one of the biggest uncertainties in eclipse calculations, the
choice of a value called $\Delta$T to predict small changes in the speed of
Earth's rotation, does not apply to this study: $\Delta$T affects the longitude
of an eclipse, but not the latitude.{[}11{]}

Interesting areas for future work would be to look more deeply at how
the frequency of solar eclipses evolves over many millennia, for
example, using Fourier analysis to study long-term eclipse cycles, and
to include Moon and Earth topology.

\section{Notes}\label{notes}

a In accordance with what was then the custom in astronomy, Meeus
treated western longitudes as positive, and eastern longitudes as
negative.{[}12{]}

b In our data, the 1st millennium has a span of 999 years, from 0001 to
0999 CE; the other 14 millennia have spans of 1,000 years.

c Meeus chose his 408 standard points as follows: ``the points at
latitudes +80°, +70°, +60°, etc., to -80° on the 24 meridians of
longitudes +180°, +165°, +150°, etc., to -165°.''

\section{Acknowledgements}\label{acknowledgements}

We are grateful to the anonymous reviewer for their exceptionally
insightful comments, suggestions and references.

\section{References}\label{references}

1 Meeus J., \emph{J. Brit. Astron. Assoc.}, 92, 124 (1982)

2 Russell H. N., Dugan R. S. and Stewart J. Q., \emph{Astronomy}, I,
227, Boston (1926)

3 Meeus J., \emph{J. R. Astron. Soc. Canada}, 74, 291 (1980)

4 Lynch T. R., \emph{Hewlett-Packard Journal}, 31, 7, 3 (1980)

5 timeanddate.com, ``Annular Solar Eclipse on 18 Apr 1977: Path Map \&
Times'' with the point 10$^\circ$S, 30$^\circ$E selected:\\
\url{https://www.timeanddate.com/eclipse/map/1977-april-18?n=%40-10%2C30}
(accessed 2025 April 18)

6 Amos, J., ``Solar eclipse 2024: Millions in North America will view
what promises to be a blockbuster'', BBC News, 2024 March 22:\\
\url{https://www.bbc.com/news/science-environment-68597945}
(accessed 2025 May 1)

7 Meeus, J., \emph{Mathematical Astronomy Morsels V}, Willmann-Bell
(2010)

8 Meeus, J., \emph{Mathematical Astronomy Morsels III}, Willmann-Bell
(2004)

9 van den Heuvel E. P. J., \emph{Geophysical Journal International}, 11,
3, 323 (1966)

10 timeanddate.com, ``Eclipses in London, England, United Kingdom'':\\
\url{https://www.timeanddate.com/eclipse/in/uk/london}
(accessed 2025 May 5)

11 Espenak, F., \emph{Fifty Year Canon of Solar Eclipses: 1986-2035},
1178, NASA Scientific and Technical Information Office (1987)

12 Wilkins, G. A., ed., \emph{Explanatory Supplement to the Astronomical
Ephemeris and the American Ephemeris and Nautical Almanac}, HM
Stationery Office (1961)

\section{Appendix}\label{appendix}

Table~A1 contains our data for the period 2000 January~1 to 2999
December~31. An explanation of the column headings is included in the
results section above. A copy of our data for the entire period from
0001 January~1 to 14999 December~31 is available for download at
\url{https://github.com/timeanddate/solar-eclipse-frequency-data}.

\setcounter{table}{0}
\renewcommand{\thetable}{A\arabic{table}}

{%
\scriptsize
\setlength{\tabcolsep}{3pt}
\renewcommand{\arraystretch}{0.95}

\begin{longtable}{
  S[table-format=2.1]
  S[table-format=1.6]
  S[table-format=4.0]
  S[table-format=3.2]
  S[table-format=3.0]
  S[table-format=1.2]
  S[table-format=3.0]
  S[table-format=1.2]
}
\caption{Our computer output for 2000--2999 CE.}\label{tab:latitude-bands}\\

\toprule
{Band midpoint} & {Area} &
{\shortstack{Count\\All}} & {\shortstack{Average\\All}} &
{\shortstack{Count\\Totality}} & {\shortstack{Average\\Totality}} &
{\shortstack{Count\\Annularity}} & {\shortstack{Average\\Annularity}} \\
\midrule
\endfirsthead

\toprule
{Band midpoint} & {Area} &
{\shortstack{Count\\All}} & {\shortstack{Average\\All}} &
{\shortstack{Count\\Totality}} & {\shortstack{Average\\Totality}} &
{\shortstack{Count\\Annularity}} & {\shortstack{Average\\Annularity}} \\
\midrule
\endhead

\bottomrule
\endfoot

-89.5 & 0.000077 & 451 & 435.79 & 6 & 3.36 & 10 & 7.24 \\
-88.5 & 0.000230 & 473 & 435.83 & 9 & 3.06 & 14 & 6.97 \\
-87.5 & 0.000384 & 488 & 436.84 & 12 & 3.04 & 17 & 6.98 \\
-86.5 & 0.000538 & 510 & 437.43 & 16 & 3.05 & 21 & 6.80 \\
-85.5 & 0.000691 & 523 & 438.22 & 18 & 3.10 & 25 & 7.14 \\
-84.5 & 0.000844 & 541 & 437.78 & 20 & 3.08 & 27 & 7.02 \\
-83.5 & 0.000997 & 559 & 437.86 & 26 & 3.02 & 32 & 7.20 \\
-82.5 & 0.001149 & 580 & 437.72 & 29 & 3.25 & 34 & 7.48 \\
-81.5 & 0.001301 & 604 & 437.75 & 33 & 3.19 & 42 & 7.34 \\
-80.5 & 0.001453 & 622 & 437.32 & 39 & 3.02 & 47 & 7.24 \\
-79.5 & 0.001604 & 643 & 437.15 & 41 & 3.24 & 51 & 7.13 \\
-78.5 & 0.001755 & 666 & 437.19 & 43 & 3.07 & 56 & 7.22 \\
-77.5 & 0.001905 & 688 & 437.35 & 48 & 2.99 & 62 & 7.45 \\
-76.5 & 0.002054 & 714 & 437.58 & 51 & 3.01 & 66 & 7.51 \\
-75.5 & 0.002203 & 737 & 437.70 & 55 & 2.80 & 71 & 7.55 \\
-74.5 & 0.002351 & 759 & 438.32 & 58 & 2.69 & 79 & 7.64 \\
-73.5 & 0.002498 & 788 & 438.66 & 63 & 2.66 & 89 & 7.70 \\
-72.5 & 0.002644 & 814 & 439.37 & 70 & 2.71 & 96 & 7.78 \\
-71.5 & 0.002790 & 841 & 440.09 & 73 & 2.70 & 105 & 7.72 \\
-70.5 & 0.002935 & 876 & 441.23 & 75 & 2.52 & 114 & 7.65 \\
-69.5 & 0.003078 & 913 & 443.04 & 80 & 2.56 & 123 & 7.80 \\
-68.5 & 0.003221 & 961 & 445.56 & 83 & 2.78 & 129 & 7.82 \\
-67.5 & 0.003363 & 1036 & 450.27 & 84 & 2.75 & 135 & 7.67 \\
-66.5 & 0.003503 & 1044 & 452.97 & 89 & 2.69 & 140 & 7.62 \\
-65.5 & 0.003643 & 1054 & 449.31 & 90 & 2.82 & 147 & 7.64 \\
-64.5 & 0.003781 & 1066 & 444.11 & 94 & 2.76 & 154 & 7.62 \\
-63.5 & 0.003918 & 1078 & 440.02 & 98 & 2.68 & 157 & 7.52 \\
-62.5 & 0.004054 & 1085 & 436.41 & 101 & 2.54 & 162 & 7.16 \\
-61.5 & 0.004189 & 1097 & 432.90 & 107 & 2.47 & 163 & 7.09 \\
-60.5 & 0.004322 & 1112 & 429.75 & 105 & 2.52 & 168 & 6.76 \\
-59.5 & 0.004453 & 1117 & 426.93 & 110 & 2.33 & 174 & 6.54 \\
-58.5 & 0.004584 & 1125 & 424.15 & 114 & 2.41 & 179 & 6.41 \\
-57.5 & 0.004713 & 1137 & 421.62 & 119 & 2.52 & 182 & 6.25 \\
-56.5 & 0.004840 & 1142 & 419.25 & 123 & 2.63 & 189 & 6.14 \\
-55.5 & 0.004966 & 1146 & 416.79 & 126 & 2.60 & 191 & 6.09 \\
-54.5 & 0.005090 & 1154 & 414.81 & 131 & 2.67 & 194 & 5.86 \\
-53.5 & 0.005213 & 1157 & 412.95 & 136 & 2.75 & 197 & 5.63 \\
-52.5 & 0.005333 & 1162 & 410.80 & 138 & 2.77 & 198 & 5.40 \\
-51.5 & 0.005453 & 1169 & 408.89 & 143 & 2.76 & 201 & 5.33 \\
-50.5 & 0.005570 & 1181 & 407.31 & 148 & 2.72 & 209 & 5.35 \\
-49.5 & 0.005686 & 1188 & 405.82 & 150 & 2.69 & 212 & 5.39 \\
-48.5 & 0.005800 & 1199 & 404.16 & 155 & 2.67 & 222 & 5.20 \\
-47.5 & 0.005912 & 1204 & 402.71 & 152 & 2.64 & 223 & 5.27 \\
-46.5 & 0.006022 & 1203 & 401.19 & 156 & 2.57 & 227 & 5.16 \\
-45.5 & 0.006131 & 1204 & 399.61 & 161 & 2.57 & 238 & 5.18 \\
-44.5 & 0.006237 & 1210 & 398.05 & 160 & 2.64 & 243 & 5.12 \\
-43.5 & 0.006342 & 1218 & 396.46 & 164 & 2.56 & 241 & 5.21 \\
-42.5 & 0.006444 & 1227 & 394.93 & 166 & 2.61 & 246 & 5.12 \\
-41.5 & 0.006545 & 1231 & 393.41 & 167 & 2.62 & 253 & 4.98 \\
-40.5 & 0.006644 & 1238 & 391.49 & 171 & 2.56 & 250 & 4.88 \\
-39.5 & 0.006740 & 1242 & 389.74 & 172 & 2.41 & 253 & 4.81 \\
-38.5 & 0.006834 & 1244 & 388.08 & 172 & 2.37 & 255 & 4.54 \\
-37.5 & 0.006927 & 1245 & 386.46 & 173 & 2.36 & 251 & 4.47 \\
-36.5 & 0.007017 & 1252 & 385.07 & 175 & 2.38 & 257 & 4.49 \\
-35.5 & 0.007105 & 1259 & 383.53 & 178 & 2.38 & 255 & 4.62 \\
-34.5 & 0.007190 & 1272 & 381.85 & 178 & 2.39 & 254 & 4.72 \\
-33.5 & 0.007274 & 1274 & 380.43 & 184 & 2.55 & 257 & 4.75 \\
-32.5 & 0.007355 & 1275 & 379.00 & 198 & 2.59 & 252 & 4.62 \\
-31.5 & 0.007434 & 1274 & 377.54 & 203 & 2.62 & 255 & 4.41 \\
-30.5 & 0.007511 & 1277 & 376.05 & 209 & 2.69 & 253 & 4.36 \\
-29.5 & 0.007586 & 1282 & 374.51 & 214 & 2.65 & 246 & 4.43 \\
-28.5 & 0.007658 & 1289 & 373.27 & 220 & 2.61 & 247 & 4.32 \\
-27.5 & 0.007728 & 1291 & 371.85 & 226 & 2.75 & 247 & 4.35 \\
-26.5 & 0.007796 & 1290 & 370.31 & 230 & 2.71 & 244 & 4.23 \\
-25.5 & 0.007861 & 1294 & 369.06 & 233 & 2.72 & 242 & 4.28 \\
-24.5 & 0.007924 & 1304 & 368.02 & 236 & 2.68 & 247 & 4.38 \\
-23.5 & 0.007984 & 1305 & 366.97 & 235 & 2.55 & 258 & 4.38 \\
-22.5 & 0.008042 & 1312 & 366.09 & 234 & 2.55 & 257 & 4.23 \\
-21.5 & 0.008098 & 1315 & 365.57 & 233 & 2.54 & 256 & 4.07 \\
-20.5 & 0.008151 & 1317 & 365.23 & 235 & 2.54 & 259 & 4.03 \\
-19.5 & 0.008201 & 1322 & 364.54 & 244 & 2.56 & 259 & 3.99 \\
-18.5 & 0.008250 & 1329 & 363.73 & 251 & 2.64 & 263 & 3.90 \\
-17.5 & 0.008295 & 1336 & 363.15 & 252 & 2.48 & 264 & 4.13 \\
-16.5 & 0.008339 & 1339 & 362.32 & 250 & 2.42 & 266 & 4.24 \\
-15.5 & 0.008380 & 1343 & 361.94 & 249 & 2.46 & 273 & 3.98 \\
-14.5 & 0.008418 & 1350 & 361.47 & 248 & 2.49 & 269 & 3.81 \\
-13.5 & 0.008454 & 1347 & 360.87 & 252 & 2.44 & 267 & 3.78 \\
-12.5 & 0.008487 & 1344 & 360.33 & 249 & 2.36 & 266 & 3.80 \\
-11.5 & 0.008518 & 1353 & 359.63 & 252 & 2.34 & 274 & 3.80 \\
-10.5 & 0.008546 & 1359 & 358.86 & 258 & 2.46 & 281 & 3.86 \\
-9.5 & 0.008572 & 1360 & 358.55 & 259 & 2.57 & 276 & 3.93 \\
-8.5 & 0.008595 & 1359 & 358.63 & 256 & 2.54 & 283 & 3.87 \\
-7.5 & 0.008615 & 1358 & 358.53 & 250 & 2.53 & 281 & 3.85 \\
-6.5 & 0.008633 & 1357 & 358.47 & 255 & 2.47 & 280 & 3.79 \\
-5.5 & 0.008649 & 1356 & 358.57 & 263 & 2.69 & 275 & 3.73 \\
-4.5 & 0.008661 & 1352 & 358.60 & 264 & 2.60 & 279 & 3.76 \\
-3.5 & 0.008672 & 1360 & 358.65 & 263 & 2.60 & 283 & 3.93 \\
-2.5 & 0.008679 & 1363 & 358.64 & 262 & 2.69 & 283 & 3.98 \\
-1.5 & 0.008685 & 1361 & 358.49 & 260 & 2.44 & 284 & 3.95 \\
-0.5 & 0.008687 & 1362 & 358.71 & 255 & 2.46 & 282 & 3.72 \\
0.5 & 0.008687 & 1361 & 358.60 & 254 & 2.51 & 287 & 3.73 \\
1.5 & 0.008685 & 1363 & 358.48 & 246 & 2.45 & 288 & 3.85 \\
2.5 & 0.008679 & 1365 & 358.48 & 245 & 2.35 & 289 & 4.01 \\
3.5 & 0.008672 & 1365 & 358.89 & 246 & 2.31 & 291 & 4.05 \\
4.5 & 0.008661 & 1363 & 359.12 & 245 & 2.33 & 289 & 3.96 \\
5.5 & 0.008649 & 1357 & 359.54 & 248 & 2.33 & 287 & 3.87 \\
6.5 & 0.008633 & 1353 & 360.10 & 245 & 2.51 & 284 & 4.00 \\
7.5 & 0.008615 & 1349 & 360.22 & 250 & 2.49 & 286 & 4.16 \\
8.5 & 0.008595 & 1349 & 360.44 & 248 & 2.60 & 288 & 4.23 \\
9.5 & 0.008572 & 1348 & 360.58 & 251 & 2.65 & 286 & 4.18 \\
10.5 & 0.008546 & 1339 & 360.95 & 247 & 2.63 & 284 & 4.10 \\
11.5 & 0.008518 & 1339 & 361.34 & 247 & 2.70 & 281 & 4.03 \\
12.5 & 0.008487 & 1336 & 361.51 & 244 & 2.79 & 277 & 4.07 \\
13.5 & 0.008454 & 1334 & 361.47 & 242 & 2.77 & 281 & 4.23 \\
14.5 & 0.008418 & 1329 & 361.56 & 239 & 2.69 & 282 & 4.25 \\
15.5 & 0.008380 & 1331 & 361.79 & 236 & 2.80 & 281 & 4.26 \\
16.5 & 0.008339 & 1328 & 362.58 & 236 & 2.82 & 279 & 4.16 \\
17.5 & 0.008295 & 1332 & 363.21 & 233 & 2.92 & 270 & 4.06 \\
18.5 & 0.008250 & 1333 & 363.77 & 234 & 2.85 & 274 & 4.07 \\
19.5 & 0.008201 & 1329 & 364.34 & 234 & 2.87 & 271 & 4.27 \\
20.5 & 0.008151 & 1317 & 365.28 & 231 & 2.97 & 269 & 4.34 \\
21.5 & 0.008098 & 1316 & 366.65 & 231 & 3.08 & 274 & 4.29 \\
22.5 & 0.008042 & 1309 & 367.82 & 229 & 3.09 & 273 & 4.12 \\
23.5 & 0.007984 & 1314 & 368.84 & 225 & 3.08 & 272 & 4.12 \\
24.5 & 0.007924 & 1314 & 369.72 & 223 & 2.85 & 270 & 4.10 \\
25.5 & 0.007861 & 1312 & 370.56 & 221 & 2.90 & 271 & 4.27 \\
26.5 & 0.007796 & 1303 & 371.72 & 218 & 2.92 & 275 & 4.35 \\
27.5 & 0.007728 & 1299 & 372.99 & 217 & 2.96 & 270 & 4.16 \\
28.5 & 0.007658 & 1300 & 374.13 & 215 & 2.85 & 261 & 3.93 \\
29.5 & 0.007586 & 1304 & 375.09 & 216 & 2.84 & 261 & 3.95 \\
30.5 & 0.007511 & 1296 & 376.29 & 206 & 2.78 & 261 & 4.09 \\
31.5 & 0.007434 & 1294 & 377.68 & 207 & 2.83 & 261 & 4.36 \\
32.5 & 0.007355 & 1285 & 379.25 & 208 & 2.88 & 256 & 4.32 \\
33.5 & 0.007274 & 1282 & 381.05 & 205 & 3.02 & 254 & 3.95 \\
34.5 & 0.007190 & 1280 & 382.78 & 202 & 3.08 & 243 & 4.02 \\
35.5 & 0.007105 & 1281 & 384.59 & 199 & 2.95 & 246 & 4.25 \\
36.5 & 0.007017 & 1272 & 386.10 & 200 & 3.14 & 245 & 4.41 \\
37.5 & 0.006927 & 1258 & 387.85 & 196 & 3.10 & 240 & 4.45 \\
38.5 & 0.006834 & 1256 & 389.68 & 191 & 3.09 & 239 & 4.45 \\
39.5 & 0.006740 & 1254 & 391.39 & 191 & 3.25 & 236 & 4.18 \\
40.5 & 0.006644 & 1249 & 392.82 & 188 & 3.10 & 233 & 4.30 \\
41.5 & 0.006545 & 1241 & 394.43 & 183 & 2.95 & 230 & 4.42 \\
42.5 & 0.006444 & 1235 & 396.26 & 175 & 2.94 & 225 & 4.61 \\
43.5 & 0.006342 & 1227 & 398.15 & 172 & 3.11 & 221 & 4.70 \\
44.5 & 0.006237 & 1214 & 400.53 & 176 & 3.16 & 217 & 4.74 \\
45.5 & 0.006131 & 1209 & 402.86 & 173 & 3.03 & 211 & 4.43 \\
46.5 & 0.006022 & 1203 & 405.02 & 163 & 2.90 & 208 & 4.51 \\
47.5 & 0.005912 & 1193 & 407.30 & 163 & 3.01 & 203 & 4.66 \\
48.5 & 0.005800 & 1187 & 409.64 & 160 & 3.13 & 204 & 4.88 \\
49.5 & 0.005686 & 1180 & 411.72 & 157 & 2.97 & 196 & 5.03 \\
50.5 & 0.005570 & 1175 & 414.33 & 154 & 2.88 & 192 & 4.71 \\
51.5 & 0.005453 & 1168 & 416.89 & 147 & 3.11 & 181 & 4.40 \\
52.5 & 0.005333 & 1159 & 419.59 & 146 & 3.17 & 182 & 4.45 \\
53.5 & 0.005213 & 1155 & 422.02 & 141 & 3.08 & 181 & 4.54 \\
54.5 & 0.005090 & 1147 & 424.14 & 138 & 3.12 & 179 & 4.68 \\
55.5 & 0.004966 & 1134 & 426.31 & 133 & 3.13 & 171 & 4.87 \\
56.5 & 0.004840 & 1127 & 428.97 & 125 & 3.02 & 173 & 4.78 \\
57.5 & 0.004713 & 1118 & 431.64 & 122 & 3.09 & 167 & 4.59 \\
58.5 & 0.004584 & 1109 & 434.53 & 120 & 3.17 & 162 & 4.59 \\
59.5 & 0.004453 & 1103 & 437.46 & 114 & 3.24 & 159 & 4.88 \\
60.5 & 0.004322 & 1099 & 440.21 & 111 & 3.37 & 157 & 5.09 \\
61.5 & 0.004189 & 1088 & 443.41 & 108 & 3.42 & 150 & 5.25 \\
62.5 & 0.004054 & 1077 & 446.81 & 106 & 3.48 & 147 & 5.49 \\
63.5 & 0.003918 & 1069 & 451.06 & 99 & 3.62 & 143 & 5.35 \\
64.5 & 0.003781 & 1062 & 456.28 & 96 & 3.61 & 137 & 5.26 \\
65.5 & 0.003643 & 1055 & 461.84 & 92 & 3.67 & 129 & 5.25 \\
66.5 & 0.003503 & 1044 & 465.86 & 89 & 3.72 & 126 & 5.38 \\
67.5 & 0.003363 & 1035 & 463.97 & 86 & 3.61 & 122 & 5.53 \\
68.5 & 0.003221 & 970 & 459.54 & 79 & 3.55 & 115 & 5.73 \\
69.5 & 0.003078 & 921 & 457.78 & 75 & 3.42 & 109 & 5.68 \\
70.5 & 0.002935 & 883 & 456.92 & 70 & 3.43 & 100 & 5.54 \\
71.5 & 0.002790 & 852 & 456.45 & 64 & 3.70 & 93 & 5.17 \\
72.5 & 0.002644 & 822 & 455.44 & 62 & 3.64 & 87 & 5.19 \\
73.5 & 0.002498 & 793 & 455.07 & 57 & 3.59 & 80 & 5.20 \\
74.5 & 0.002351 & 771 & 454.70 & 53 & 3.79 & 76 & 5.21 \\
75.5 & 0.002203 & 745 & 454.07 & 51 & 4.11 & 67 & 5.07 \\
76.5 & 0.002054 & 720 & 453.42 & 49 & 4.10 & 58 & 5.34 \\
77.5 & 0.001905 & 695 & 452.98 & 46 & 4.05 & 56 & 5.48 \\
78.5 & 0.001755 & 669 & 452.09 & 43 & 4.08 & 53 & 5.51 \\
79.5 & 0.001604 & 644 & 451.50 & 38 & 3.99 & 47 & 5.71 \\
80.5 & 0.001453 & 625 & 451.71 & 37 & 3.84 & 44 & 5.67 \\
81.5 & 0.001301 & 605 & 451.80 & 33 & 4.19 & 43 & 5.43 \\
82.5 & 0.001149 & 589 & 451.08 & 31 & 4.75 & 39 & 4.97 \\
83.5 & 0.000997 & 566 & 450.86 & 31 & 5.06 & 28 & 4.95 \\
84.5 & 0.000844 & 551 & 449.97 & 28 & 4.60 & 26 & 5.29 \\
85.5 & 0.000691 & 533 & 448.54 & 21 & 3.74 & 24 & 5.86 \\
86.5 & 0.000538 & 511 & 447.52 & 16 & 3.04 & 23 & 6.03 \\
87.5 & 0.000384 & 496 & 446.13 & 11 & 3.23 & 19 & 5.77 \\
88.5 & 0.000230 & 478 & 446.35 & 9 & 3.97 & 13 & 5.75 \\
89.5 & 0.000077 & 464 & 445.02 & 8 & 3.84 & 9 & 5.97 \\

\end{longtable}
}

\end{document}